\documentclass[12pt,twoside]{article}
\usepackage{fleqn,espcrc1,psfig,epsf}

\usepackage{graphicx}
\usepackage[figuresright]{rotating}

\newcommand{\AmS}{{\protect\the\textfont2
  A\kern-.1667em\lower.5ex\hbox{M}\kern-.125emS}}

\title{Flavor asymmetry of the nucleon sea}

\author{J. C. Peng\address{Physics Division,
        Los Alamos National Laboratory, \\ 
        Los Alamos, New Mexico, 87545 U.S.A.}}
       
\begin{document}

\maketitle

\begin{abstract}
Recent deep inelastic scattering and Drell-Yan experiments have revealed 
a surprisingly large asymmetry between the up and down sea quark
distributions in the nucleon. 
The current status of the flavor asymmetry of the nucleon sea is reviewed. 
Implications of various theoretical models and possible 
future measurements are also discussed.
\end{abstract}

\section{INTRODUCTION}

One of the most active areas of research in nuclear and particle physics 
during the last several decades is the study of quark and gluon distributions 
in the nucleons and nuclei. Several major surprises were discovered in 
Deep-Inelastic Scattering (DIS) experiments which profoundly changed our 
views of the partonic substructure of hadrons. In the early 1980's, the 
famous `EMC' effect found in muon DIS provided the first 
unambiguous evidence that the quark distributions in nuclei are
significantly different from those in free nucleons~\cite{emc83,gee95}. 
More recently, surprising results on the spin and
flavor structures of the nucleons were discovered in DIS experiments. 
The so-called ``spin crisis'', revealed by the disagreement between the 
prediction of the Ellis-Jaffe sum rule and the polarized DIS 
experiments, has led to extensive theoretical and experimental efforts to 
understand the partonic content of proton's 
spin~\cite{hughes99}. Subsequently, the observation~\cite{nmc91} of the 
violation of the Gottfried sum rule~\cite{gott} in DIS revealed a 
surprisingly large asymmetry between the up and down antiquark distributions 
in the nucleon, shedding new light on the origins of the nucleon sea.

In this article, we review the status of our current knowledge on the
flavor dependence of the sea quark distributions in hadrons. 
In Section 2, we review the early studies
of the nucleon sea in DIS and lepton-pair production. The crucial recent 
experiments establishing the up/down flavor asymmetry of the nucleon sea
are also discussed in Section 2. Various theoretical models for 
explaining the $\bar d/ \bar u$ asymmetry are described in Section 3.
The implications of these models on other aspects of the
parton structure functions are discussed in Section 4. Finally, we
present future prospects in Section 5, followed by 
conclusion in Section 6.

\section{EXPERIMENTAL EVIDENCE FOR $\bar d / \bar u$ ASYMMETRY}

The earliest parton models assumed that the proton sea was flavor symmetric,
even though the valence quark distributions are clearly flavor asymmetric.
Inherent in this assumption is that the content of the sea is 
independent of the valence quark's composition. Therefore,
the proton and neutron were expected to have identical sea-quark distributions.
The flavor symmetry assumption was not based on any known physics, and 
it remained to be tested by experiments. Neutrino-induced charm production
experiments~\cite{abrom82,conrad98}, which are 
sensitive to the $s \to c$ process, provided strong
evidences that the strange-quark content of the nucleon is only about half
of the up or down sea quarks. Such flavor asymmetry is attributed to the
much heavier strange-quark mass compared to the up and down quarks. The similar
masses for the up and down quarks suggest that the nucleon sea should be nearly
up-down symmetric. 

The issue of the equality of $\bar u$ and $\bar d$ was first
encountered in measurements of the Gottfried integral~\cite{gott},
defined as
\begin{equation}
I_G = \int_0^1 \left[F^p_2 (x,Q^2) - F^n_2 (x,Q^2)\right]/x~ dx,
\end{equation}
where $F^p_2$ and $F^n_2$ are the proton and neutron structure
functions measured in DIS
experiments. Assuming charge symmetry, 
$I_G$ can be expressed in terms of the valence and sea quark
distributions of the proton as:
\begin{eqnarray}
I_G  = \frac{1}{3} \int_0^1 \left[u_v (x, Q^2) - d_v (x,Q^2)\right] 
dx + \frac{2}{3} \int_0^1 \left[\bar u (x,Q^2) - \bar d (x,Q^2)\right] dx.
\label{eq:ig}
\end{eqnarray}
Under the assumption of a $\bar u$, $\bar d$ flavor-symmetric sea in
the nucleon, the Gottfried Sum Rule (GSR)~\cite{gott}, $I_G
= 1/3$, is obtained.

The first measurements of the Gottfried integral were carried out
in the 1970s~\cite{bloom70,bloom73} soon after the discovery of scaling in DIS. 
Although the large systematic errors associated with 
the unmeasured small-$x$ region prevented a sensitive test of the GSR, 
Field and Feynman~\cite{field77} nevertheless interpreted the early 
SLAC data as strong indications that GSR is violated and that the 
$\bar u$ and $\bar d$ distributions in the proton are different.
The SLAC DIS experiments were followed by several muon-induced DIS experiments
at Fermilab and at CERN. Despite the fact that all measurements of 
Gottfried integral consistently showed
a value lower than 1/3, the large systematic errors prevented a
definitive conclusion. As a result, all 
parametrizations~\cite{do84,ehlq84,abfow89}
of the parton distributions based on global fits to existing data before 1990 
assumed a symmetric $\bar u$, $\bar d$
sea. 

The most accurate test of the GSR was reported in 1991 by the New Muon 
Collaboration (NMC)~\cite{nmc91}, which measured $F^p_2$ and $F^n_2$ over the 
region $0.004 \le x \le 0.8$. They determined the Gottfried integral to be 
$ 0.235\pm 0.026$, significantly below 1/3. This surprising result has
generated much interest.

Although the violation of the GSR can be explained by
assuming unusual behavior of the parton distributions at very
small $x$,
a more natural explanation is to abandon the assumption
$\bar u = \bar d$.  Specifically, the NMC result implies

\begin{equation}
\int_0^1 \left[\bar d(x) - \bar u(x)\right] dx = 0.148 \pm 0.039.
\label{eq:3.2}
\end{equation}
It should be emphasized that only the integral of $\bar d -\bar u$ was deduced
from the DIS measurements. The $x$ dependence of $\bar d - \bar u$ remained
unspecified.
\begin{figure}
\center
\hspace*{0.05in}
\psfig{figure=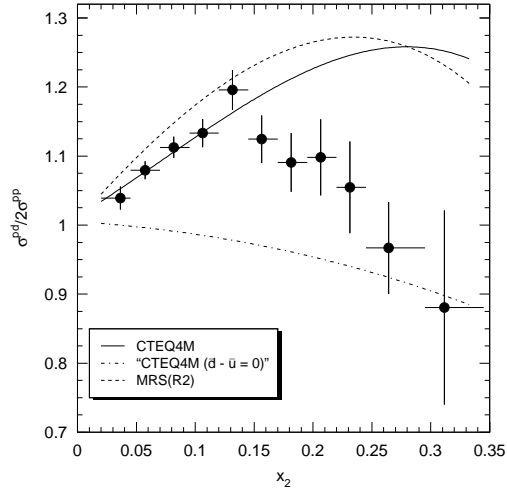,height=2.6in}
\vspace*{-0.3in}
\caption{The ratio $\sigma^{pd}/2\sigma^{pp}$ of Drell-Yan cross
sections {\em vs.} $x_{2}$ from Fermilab E866~\cite{e866}. 
The curves are next-to-leading order
calculations, weighted by acceptance, of the Drell-Yan cross section 
ratio using the CTEQ4M~\cite{cteq} and MRS(R2)~\cite{mrs} parton distributions. 
In the lower
CTEQ4M curve, $\bar{d} - \bar{u}$ was set to 0 to simulate a symmetric sea.}
\label{fig:3.1}
\vspace*{-0.25in}
\end{figure}

The proton-induced Drell-Yan (DY) process provides an
independent means to probe the flavor asymmetry of the nucleon sea~\cite{es}.
An important advantage of the DY process is that the $x$ dependence of 
$\bar d / \bar u$ can be determined.
It is interesting to note that, as early as 1981, Fermilab E288~\cite{ito} 
reported evidence for a $\bar d/\bar u$ asymmetry, 
based on a measurement of the $p + d$ DY cross
section. However, this interpretation depended sensitively on assumptions
about the shape of the valence quark distributions and was 
not conclusive. Later, the 
Fermilab E772 collaboration~\cite{plm91}
compared the DY yields from isoscalar targets with that from a neutron-rich 
(tungsten) target, and constraints on the nonequality of $\bar u$ and 
$\bar d$ in the range $0.04 \leq x \leq 0.27$ were set. 
More recently, the CERN experiment NA51~\cite{na51}
carried out a comparison of the DY muon pair yield from hydrogen and
deuterium using a 450-GeV/c proton beam. They found that
$\bar u / \bar d = 0.51 \pm 0.04 \pm 0.05$ at $\langle x \rangle =0.18$, a 
surprisingly large difference between the $\bar u$ and $\bar d$.

A DY experiment (E866), aiming at higher statistical accuracy and wider 
kinematic coverage than NA51, was recently completed~\cite{e866,peng}
at Fermilab. This experiment also measured the 
DY muon pairs from 800-GeV/c protons interacting with liquid deuterium and hydrogen targets.
The acceptance of the spectrometer
was largest for $x_F = x_1 - x_2 > 0$. In this kinematic regime the DY
cross section is dominated by the annihilation of a beam quark with a target
antiquark. The DY cross section ratio
at large $x_F$ is approximately given as 
\begin{equation}
{\sigma_{DY}(p+d)\over 2\sigma_{DY}(p+p)} \approx
{1\over 2} \left(1+{\bar d(x_2)\over \bar u(x_2)}\right).
\label{eq:3.3}
\end{equation}
The ratio is unity when  $\bar d = \bar u$.
Figure~\ref{fig:3.1} shows that the E866 measurement of this ratio 
clearly exceeds unity for an appreciable range in $x_2$.

\begin{figure}
\center
\hspace*{-0.2in}
\psfig{figure=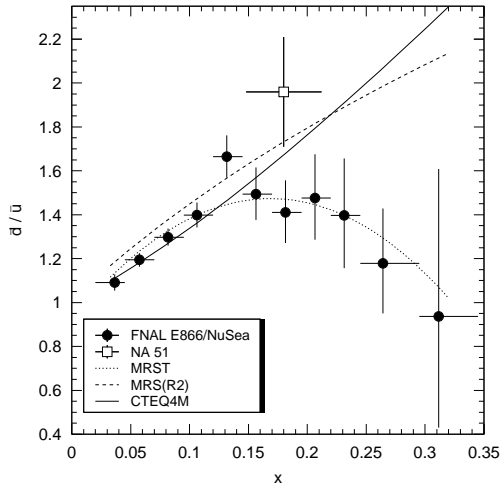,height=2.60in}
\vspace*{-0.35in}
  \caption{The ratio of $\bar{d}/\bar{u}$ in the proton as a function
  of $x$ extracted from the Fermilab E866~\cite{e866} cross section ratio. The
  curves are from various parton distributions.  
  Also shown is the result
  from NA51~\cite{na51}, plotted as an open box.}
\label{fig:3.2}
\vspace*{-0.25in}
\end{figure}

Using an iterative procedure~\cite{peng}, values for 
$\bar d/ \bar u$ were extracted by the E866 collaboration at
$Q^2 = 54$ GeV$^2$/c$^2$. These are shown in Figure~\ref{fig:3.2} along 
with the NA51 measurement.
For $x < 0.15$, $\bar d/\bar u$ increases linearly with $x$ and is in
good agreement with the CTEQ4M~\cite{cteq} and MRS(R2)~\cite{mrs} parameterizations. 
However, a distinct feature of the data, not seen in either 
parameterization, is the
rapid decrease toward unity of $\bar{d}/\bar{u}$ beyond
$x=0.2$\@. 

\begin{figure}
\center
\hspace*{0.05in}
\psfig{figure=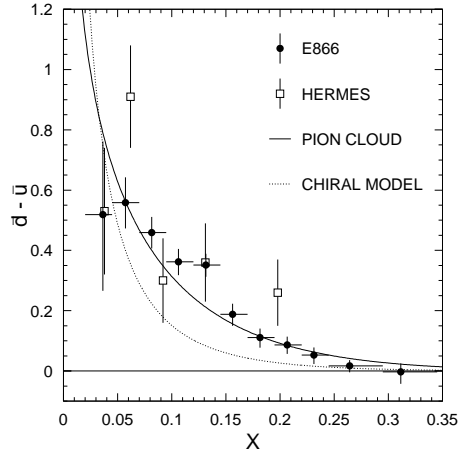,height=2.60in}
\vspace*{-0.35in}
\caption{Comparison of the E866~\cite{e866} $\bar d - \bar u$ results at $Q^2$ =
54 GeV$^2$/c$^2$ with the predictions of pion-cloud and chiral models 
as described in the text. The data from HERMES~\cite{hermes} are also shown.}
\label{fig:3.3}
\vspace*{-0.25in}
\end{figure}

The $\bar d / \bar u$ ratio, along with the
CTEQ4M values for $\bar d + \bar u$, was used to obtain
$\bar d - \bar u$ over the region $0.02 < x < 0.345$
(Figure~\ref{fig:3.3}). 
From the results shown in Figure~\ref{fig:3.3}, one can
obtain an independent determination~\cite{peng} of the integral of 
Equation~\ref{eq:3.2}.
E866 finds $0.100 \pm 0.007
\pm 0.017$, consistent with, but roughly $2/3$ of the value deduced by NMC. 

The HERMES collaboration recently reported a semi-inclusive
DIS measurement of charged pions from hydrogen and deuterium 
targets~\cite{hermes}.
Based on the differences between charged-pion yields from the two targets,
the ratio $(\bar d - \bar u)/ (u -d)$ is determined
in the kinematic range, $0.02 < x < 0.3$ and 
1 GeV$^2$/c$^2 < Q^2 <$ 10 GeV$^2$/c$^2$. The HERMES results for
$\bar d - \bar u$, shown in Figure~\ref{fig:3.3}, are consistent with
the E866 results obtained at significantly higher $Q^2$.

\section{ORIGINS OF THE $\bar d / \bar u$ ASYMMETRY}

The earliest experiments indicated that the value of the Gottfried
integral might be less than 1/3, leading to speculation regarding the origin
of this reduction. Field and Feynman suggested~\cite{field77} 
that it could be due to Pauli blocking in so far as $u \bar u$ pairs 
would be suppressed relative to $d \bar d$ pairs
because of the presence of two $u$-quarks in proton as compared to a single
$d$-quark.  Ross and Sachrajda~\cite{ross79} 
questioned that this effect would be
appreciable because of the large phase-space available to the created
$q \bar q$ pairs. They also showed that perturbative QCD would not 
produce a $\bar d / \bar u$ asymmetry. Steffens and Thomas~\cite{stef97} 
recently looked into this issue,
explicitly examining the consequences of  Pauli blocking. They
similarly concluded that the blocking effects were small.

A natural origin for this flavor asymmetry is the virtual states of
the proton containing isovector mesons. This point appears to have first
been made by Thomas in a publication ~\cite{thomas83} 
treating SU(3) symmetry breaking in
the nucleon sea. Sullivan~\cite{sullivan} previously showed that in DIS 
virtual mesons scale in the Bjorken limit and contribute to the
nucleon structure function.  Following the publication of the NMC
result, many 
papers treated virtual mesons as the origin of the $\bar d/\bar u$ asymmetry
(see~\cite{kumano0,peng99} for recent reviews).
Here the $\pi^+(\bar d u)$ cloud, dominant in the 
process $p\rightarrow\pi^+ n$, leads to an excess of $\bar d$ sea.

A different approach for including the effects of virtual mesons has
been presented by Eichten et al.~\cite{ehq} and
further investigated by other authors~\cite{cheng1,szczurek2}. In 
chiral perturbation theory, the relevant degrees of
freedom are constituent quarks, gluons, and Goldstone bosons. In
this model, a portion of the sea comes from the couplings of Goldstone
bosons to the constituent quarks, such as $u \to d \pi^+$ and $d \to u
\pi^-$. The excess of $\bar d$ over $\bar u$ is then simply due to the
additional valence $u$ quark in the proton. 

The $x$ dependences of $\bar d - \bar u$ and $\bar d / \bar u$ obtained
by E866 provide important constraints for theoretical models.
Figure~\ref{fig:3.3} compares $\bar d(x) - \bar u(x)$ from E866
with a virtual-pion model calculation, following the procedure detailed
by Kumano~\cite{kumano}. 
Figure~\ref{fig:3.3} ({\it dotted curve}) also shows the predicted
$\bar d - \bar u$ from the chiral model.
We follow the
formulation of Szczurek et al~\cite{szczurek2} to calculate
$\bar d(x) - \bar u(x)$ at $Q^2$ = 0.25 GeV$^2$/c$^2$ and then 
evolve the results
to $Q^2$ = 54 GeV$^2$/c$^2$. The chiral model places 
more strength at low $x$
than does the virtual-pion
model. This difference reflects the fact
that the pions are softer in the chiral model, since they are coupled
to constituent quarks, that carry only a fraction of the
nucleon momentum. The $x$ dependence of the E866 data favors the
virtual-pion model over the chiral model, suggesting that
correlations between the chiral constituents should be taken into
account.

Recently, the flavor asymmetry of the nucelon sea was computed 
in the large-$N_c$ limit, where the nucleon is described as a 
soliton of an effective chiral theory~\cite{waka98,poby99}.
In this chiral quark-soliton model, the flavor non-singlet 
distribution, $\bar d(x) - \bar u(x)$, appears in the next-to-leading
order of the $1/N_c$ expansion~\cite{diak96,diak97}. The E866 
$\bar d(x) - \bar u(x)$ data were shown to be well described by 
this model~\cite{poby99}.

Instantons have been known as theoretical constructs since the
seventies~\cite{bel75,hooft76,shu98}. 
The collision between a quark and an
instanton flips the helicity of the quark while creating a $q \bar q$
pair of different flavor. Thus, interaction between a $u$ quark and 
an instanton results in a $u$ quark of opposite helicity and 
either a $d \bar d$ or $s \bar s$ pair. 
Such a model has the possibility of accounting for both the flavor
asymmetry and the ``spin crisis"~\cite{forte89}. 
However, the prediction~\cite{inst93} at large 
$x$, $\bar d(x) / \bar u(x) \to 4$, 
is grossly violated by experiment (see Figure~\ref{fig:3.2}). 
Thus, it appears that while instantons have the
possibility for accounting for flavor and spin anomalies, the approach is not
yet sufficiently developed for a direct comparison.

\section{IMPLICATIONS OF THE MESON-CLOUD MODELS}

Models in which virtual mesons are admitted as degrees of freedom
have implications that extend beyond the $\bar d / \bar u$ 
flavor asymmetry addressed above.
They create hidden strangeness in the nucleon via such virtual processes as 
$p \to \Lambda + K^+, \Sigma + K$, etc. 
Such processes are of considerable interest as they imply different $s$ and
$\bar s$ parton distributions in the nucleon, a feature not found in gluonic
production of $s \bar s$ pairs. This subject has 
received a fair amount of attention
in the literature~\cite{holt96,signal87,warr92,ji95,ma96} 
but experiments have yet to clearly identify such a
difference. 

A difference between the $s$ and $\bar s$ distribution can be made manifest
by direct measurement of the $s$ and $\bar s$ parton distribution functions 
in DIS neutrino scattering, or in
the measurement of the $q^2$ dependence of the 
strange quark contribution ($F^p_{1s}(q^2)$)
to the proton charge form factor. 
This latter case follows from a suggestion of Kaplan 
and Manohar~\cite{kaplan88} regarding the new information contained in the
weak neutral current form factors of the nucleon. Measurement of these form
factors allows extraction of the strangeness contribution to the 
nucleon's charge and magnetic moment and axial form factors. The portion 
of the charge form factor $F^p_{1s} (q^2)$ due to strangeness clearly is
zero at $q^2 = 0$, but if the $s$ and $\bar s$ distributions are different the
form factor becomes non-zero at finite $q^2$. These ``strange'' form 
factors can be measured in neutrino elastic scattering~\cite{garvey93}
from the nucleon, or by selecting the parity-violating component of
electron-nucleon elastic scattering, as is now
being done at the Bates~\cite{mueller} and Jefferson Laboratories~\cite{aniol}.

It is worth pointing out that there is a relationship between the parton
distributions and the form factors of a hadron.  If the neutron's charge 
form factor is explained in terms of a particular meson-baryon expansion,
then one should expect that the expansion is consistant with the neutron''s
partonic structure.

The chiral and the meson-cloud models both predict that the $\bar u$ and
$\bar d$ quarks will carry negligible amount of the proton's 
spin~\cite{peng,ehq,cheng2}. 
In striking contrast, the chiral-quark model predicts~\cite{dress99} 
a significantly polarized
sea with a large flavor asymmetry. In particular, this model
predicts a larger values for $\Delta \bar u - \Delta \bar d$
than for $\bar d - \bar u$.
Both the SMC~\cite{smc98} and the HERMES~\cite{hermes99} 
experiments attempted to extract
the sea-quark polarizations via semi-inclusive polarized DIS measurements,
and the results indicate small sea-quark polarization consistent with
zero. However, as pointed out in Ref.~\cite{dress99}, large uncertainties
are associated with certain assumptions made in the extraction.
Future DY and $W^\pm$ production experiments at RHIC could clearly test
these models~\cite{dres99b}.

The observation
of a large $\bar d / \bar u$ asymmetry in the proton has 
motivated Alberg et al.~\cite{alberg1,alberg2}
to consider the sea-quark distributions in the $\Sigma$. The meson-cloud
model implies a $\bar d / \bar u$ asymmetry in the $\Sigma^+$ even larger than
that of the proton. However, the opposite effect is expected
from SU(3) symmetry.
Although relatively intense
$\Sigma^+$ beams have been produced for recent experiments at Fermilab,
this experiment appears to be very challenging because of
large pion, kaon, and proton contaminations in the beam.

\section{FUTURE PROSPECT}

The interplay between the perturbative and non-perturbative components of
the nucleon sea remains to be better determined. Since the perturbative
process gives a symmetric $\bar d/ \bar u$ while a non-perturbative 
process is needed to generate an asymmetric $\bar d/ \bar u$ sea, the relative
importance of these two components is directly reflected in the $\bar d/ \bar u$
ratios. Thus, it would be very important to extend the DY
measurements to kinematic regimes beyond the current limits. 

The new 120 GeV Fermilab Main Injector (FMI) and the proposed 50 GeV 
Japanese Hadron Facility~\cite{nagamiya} (JHF) present opportunities for 
extending the $\bar d/ \bar u$ measurement to larger $x$ ($x > 0.25$).
For given values of $x_1$ and $x_2$ the DY cross section
is proportional to $1/s$, hence the DY cross section 
at 50 GeV is roughly 16 times greater than
that at 800 GeV! 
The values of
$\bar d/ \bar u$ over the region $0.25 < x < 0.7$ could indeed be
obtained at FMI and JHF~\cite{p906,peng00}, and these data would be
extremely valuable for testing various theoretical models.

At the other end of the energy scale, RHIC will operate soon in the range
$50 \le \sqrt s \le 500$ GeV/nucleon. The capability of accelerating and
colliding a variety of beams from $p + p$, $p + A$, to $A + A$ at RHIC
offers a unique opportunity to extend the DY $\bar d / \bar u$ measurement
to very small $x$. Furthermore, an interesting quantity to be 
measured at RHIC is the ratio of the 
$p + p \to W^+ + X$ and $p + p \to W^- + X$ cross sections~\cite{peng1}. 
It can be shown that this
ratio is very sensitive to $\bar d / \bar u$. An important feature of
the $W$ production asymmetry in $p + p$ collision is that it is completely free 
from the assumption of charge symmetry. 

As discussed earlier, an interesting consequence of the meson-cloud model
is that the $s$ and $\bar s$ distributions
in the proton could have very different shapes, even though the net amount
of strangeness in the proton vanishes. By 
comparing the $\nu$ and $\bar \nu$
induced charm production, the CCFR 
collaboration found no difference between 
the $s$ and $\bar s$ distributions~\cite{ccfr95}. More precise
future measurements would be very helpful.
Dimuon production experiments using $K^\pm$ beams might provide an independent
determination of the $s$/$\bar s$ ratio of the proton, provided that our
current knowledge on valence-quark distributions in kaons is improved.
Ongoing measurements of $F^p_{1s}$ via 
parity-violating electron-nucleon scattering should shed much light on
the possible difference between $s$ and $\bar s$ distributions.

DY and $W^\pm$ production in polarized $p+p$ collision 
are planned at RHIC~\cite{bunce} and they have great potential for providing
qualitatively new information about antiquark polarization. At large
$x_F$ region ($x_F > 0.2$), the longitudinal spin asymmetry $A_{LL}$ in the
$\vec p + \vec p$ DY process is given by~\cite{moss,plm}
\begin{eqnarray}
A^{DY}_{LL}(x_1,x_2) \approx g_1(x_1)/F_1(x_1) \times {\Delta \bar u
\over \bar u}(x_2),
\label{eq:6.4.1}
\end{eqnarray}  
where $g_1(x)$ is the proton polarized structure function measured in DIS,
and $\Delta \bar u(x)$ is the polarized $\bar u$ distribution function.
Eq.~\ref{eq:6.4.1} shows that $\bar u$ polarization can be determined using 
polarized DY at RHIC. 

\section{CONCLUSION}

The flavor asymmetry of the nucleon sea has been clearly established
by recent DIS and DY experiments. The surprisingly large asymmetry
between $\bar u$ and $\bar d$ is unexplained by perturbative
QCD. Thus far, three distinct explanations have been offered in the 
literature for the origin of the observed $\bar d$, $\bar u$ asymmetry. 
The first is Pauli blocking which, while appealing, is difficult to calculate 
and appears to produce too small an effect to be the sole origin of 
the large observed asymmetry. 
The second involves virtual isovector mesons, mostly pions, in the nucleon. 
Such descriptions necessarily require non-perturbative QCD and, apart from 
lattice gauge calculations, demand additional parameters and possess 
systematic uncertainties. However, as these virtual mesons readily generate 
a large $\bar d / \bar u$ asymmetry, many authors, using a variety of 
techniques to invoke and justify their approachs, have obtained qualitative 
agreement with the experimental measurements. The third explanation involves 
instantons, but the theory is not sufficiently developed to allow 
quantitative comparison to the asymmetry data.
Future experiments will test and refine these models. They will further
illuminate the interplay between the perturbative and non-perturbative
nature of the nucleon sea.

\end{document}